\pdfoutput=1
\documentclass[a4paper,american,aps,citeautoscript,floatfix,longbibliography,pdftex,prl,showpacs,superscriptaddress,twocolumn]{revtex4-1}

\usepackage{amsfonts,amsmath,amssymb}
\usepackage{graphicx}
\usepackage[T1]{fontenc}
\usepackage[utf8]{inputenc}
\usepackage{microtype}
\usepackage{pxfonts,tgpagella}
\usepackage{subfigure}
\usepackage{textcomp}
\usepackage{xspace}
\usepackage{hyperref, hypernat}

\newlength{\figwidth}
\newcommand{\cost}{\ensuremath{\langle\cos^2\!\theta_\text{2D}\rangle}\xspace}
\newcommand{\costhreeD}{\ensuremath{\langle\cos^2\!\theta\rangle}\xspace}
\newcommand{\eg}{e.\,g.}%
\newcommand{\ie}{i.\,e.}%

\newcommand{\ket}[1]{\ensuremath{|#1\rangle}\xspace}%
\newcommand{\taulaser}{\ensuremath{\tau_\textup{laser}}\xspace}
\newcommand{\taurot}{\ensuremath{\tau_\textup{rot}}\xspace}

\newcommand{\cfeldesy}{\affiliation{Center for Free-Electron Laser Science, DESY, Notkestrasse 85, 22607 Hamburg, Germany}}%
\newcommand{\uhhcui}{\affiliation{The Hamburg Center for Ultrafast Imaging, University of Hamburg, Luruper Chaussee 149, 22761 Hamburg, Germany}}%
\newcommand{\uhhphys}{\affiliation{Department of Physics, University of Hamburg, Luruper Chaussee 149, 22761 Hamburg, Germany}}%
\newcommand{\granada}{\affiliation{Instituto Carlos I de F\'{\i}sica Te\'orica y Computacional and
      Departamento de F\'{\i}sica At\'omica, Molecular y Nuclear, Universidad de Granada, 18071 Granada, Spain}}%
\newcommand{\aarhuschem}{\affiliation{Department of Chemistry, Aarhus University, 8000 Aarhus C, Denmark}}%
\newcommand{\aarhusnano}{\affiliation{Interdisciplinary Nanoscience Center (iNANO), Aarhus University, 8000 Aarhus C, Denmark}}

\begin{document}
\title{Strongly driven quantum pendulum of the OCS molecule}%
\author{Sebastian Trippel}\cfeldesy%
\author{Terry Mullins}\cfeldesy%
\author{Nele L.\,M.\ Müller}\cfeldesy%
\author{Jens S.\ Kienitz}\cfeldesy\uhhcui%
\author{Juan J. Omiste}\granada%
\author{Henrik Stapelfeldt}\aarhuschem\aarhusnano%
\author{Rosario Gonz{\'a}lez-F{\'e}rez}\uhhcui\granada%
\author{\mbox{Jochen~Küpper}}\email{jochen.kuepper@cfel.de}\homepage{http://desy.cfel.de/cid/cmi}\cfeldesy\uhhcui\uhhphys%
\date{\today}%
\begin{abstract}\noindent%
   We demonstrate and analyze a strongly driven quantum pendulum in the angular motion of
   state-selected and laser aligned OCS molecules. Raman-couplings during the rising edge of a
   50-picosecond laser pulse create a wave packet of pendular states, which propagates in the
   confining potential formed by the polarizability interaction between the molecule and the laser
   field. This wave-packet dynamics manifests itself as pronounced oscillations in the degree of
   alignment with a laser-intensity dependent period.%
\end{abstract}
\maketitle%
\noindent%
Pendular states, directional superpositions of field-free rotational states, are created by the
anisotropic interaction between an isolated molecule and a strong electric
field~\cite{Loesch:JCP93:4779, Friedrich:Nature353:412, Friedrich:PRL74:4623,
   Stapelfeldt:RMP75:543}. In a classical sense, this corresponds to the free rotation of the
molecule changing into a restricted angular motion, where a molecular axis librates about the field
direction. In the case of a strong static electric field the pendular states result from the
interaction with the permanent dipole moment. This was exploited, for instance, for the
simplification of spectroscopic signatures of large molecular clusters~\cite{Nauta:Science283:1895}.
In the case of a nonresonant laser field the pendular states are formed due to the interaction with
the molecular polarizability. This interaction constitutes the basis for laser-induced
alignment~\cite{Friedrich:PRL74:4623, Stapelfeldt:RMP75:543}, the confinement of molecular axes to
laboratory-fixed axes defined by the polarization of the alignment field. Notably, in the limit
where the laser field is turned on slowly compared to the inherent rotational period(s) of the
molecule, the initial field-free rotational states are converted into the corresponding pendular
states. This process is called adiabatic alignment~\cite{Stapelfeldt:RMP75:543} and it has found
widespread use in molecular sciences~\cite{Spence:PRL92:198102, Madsen:PRL102:073007,
   Purcell:PRL103:153001, Holmegaard:NatPhys6:428, Boll:PRA88:061402, Kuepper:LCLSdibn:inprep}. The
pendular states persist for as long as the laser field is turned on and the molecules return to
their initial field-free rotational states upon turn-off of the laser field, provided this occurs
slowly compared to the rotational period(s),~$\taurot$.

Pendular states were investigated through frequency-resolved spectroscopy~\cite{Block:PRL68:1303,
   Kim:JCP104:1147} and by photodissociation or Coulomb explosion
imaging~\cite{Stapelfeldt:PRL74:3780, Stapelfeldt:RMP75:543}. The former approach probes the
field-induced changes of the rotational energy levels, thus the pendular state energies, while the
latter approach probes the way the molecules are confined in space, \ie, the orientational character
of the pendular states. So far these studies were all performed in the adiabatic limit where the
classical signature of the pendular states, \ie, the librational motion of a molecular axis about
the field direction, cannot be observed directly. To observe this motion it would be necessary to
create a coherent superposition of pendular states.

Here, we demonstrate that such pendular motion can be induced through the use of a laser pulse with
a duration $\taulaser\sim\taurot$ in between the common limits of adiabatic ($\taulaser\gg\taurot$)
and impulsive ($\taulaser\ll\taurot$) alignment. The intermediate regime has, hitherto, only been
explored theoretically~\cite{Ortigoso:JCP110:3870, Seideman:PRL83:4971, Torres:PRA72:023420}. We
performed a combined experimental and theoretical study in this intermediate pulse-duration regime
on aligned carbonyl sulfide (OCS) molecules in their absolute ground state. A wave packet of
pendular states was created by the rising edge of a 50-picosecond laser pulse and its propagation
was studied during and after the laser pulse.

\begin{figure}[t]
   \centering
   \includegraphics[width=\linewidth]{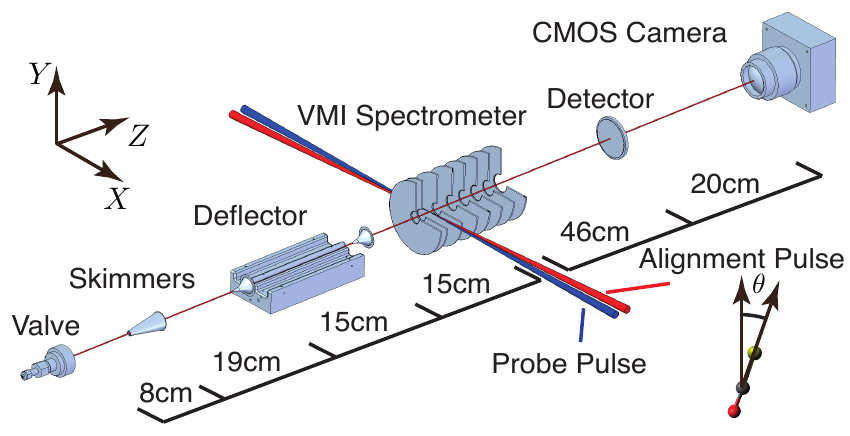}
   \caption{Schematic of the experimental setup, including the axis system and the definition of
      $\theta$ between the laboratory fixed $Y$ axis and the molecular axis $z$.}
   \label{fig:setup}
\end{figure}

A schematic of the experimental setup is shown in \autoref{fig:setup}. In short, a pulsed molecular
beam was provided by expanding 500~ppm of OCS seeded in 6~bar of neon through a cantilever piezo
valve~\cite{Irimia:RSI80:3958} at a repetition rate of 250~Hz. After passing two skimmers the
molecular beam entered the electric deflector, where the molecules were dispersed according to their
quantum state~\cite{Filsinger:JCP131:064309} and a pure sample of ground-state OCS was
selected~\cite{Nielsen:PCCP13:18971}. These molecules were aligned by a moderately intense (around
$10^{11}$~W/cm$^2$) laser pulse inside a velocity map imaging (VMI)
spectrometer~\cite{Eppink:RSI68:3477}. The angular confinement was probed through strong-field
multiple ionization by a 30~fs laser pulse resulting in Coulomb explosion of the molecule. The
produced S$^+$ ions were velocity mapped onto a 40~mm diameter position sensitive detector
consisting of a multi-channel-plate, a fast phosphor screen, and a high frame-rate camera.

The alignment and ionization laser pulses were provided by an amplified femtosecond laser system at
a repetition rate of 1~kHz and with a center wavelength of 800~nm and a spectral bandwidth of
72~nm~\cite{Trippel:MP111:1738}. Directly behind the amplification stages the laser beam was split
into two parts, an alignment beam and a probe beam. The alignment pulses can be compressed or
stretched (negatively chirped) with an external compressor continuously to pulse durations ranging
from 40~fs to 520~ps. The probe pulses were compressed to 30~fs using the standard grating based
compression setup. Since both beams were generated by the same laser system they were inherently
synchronized. Both beams were incident on a 60~cm-focal-length lens, parallel to each other, with a
transverse distance of 10~mm. The foci were overlapped in space and time in the molecular beam in
the center of the velocity-map-imaging spectrometer. The relative timing between the two pulses was
adjusted by a motorized linear translation stage.

\begin{figure}
   \centering
   \includegraphics[angle=0, width=\linewidth]{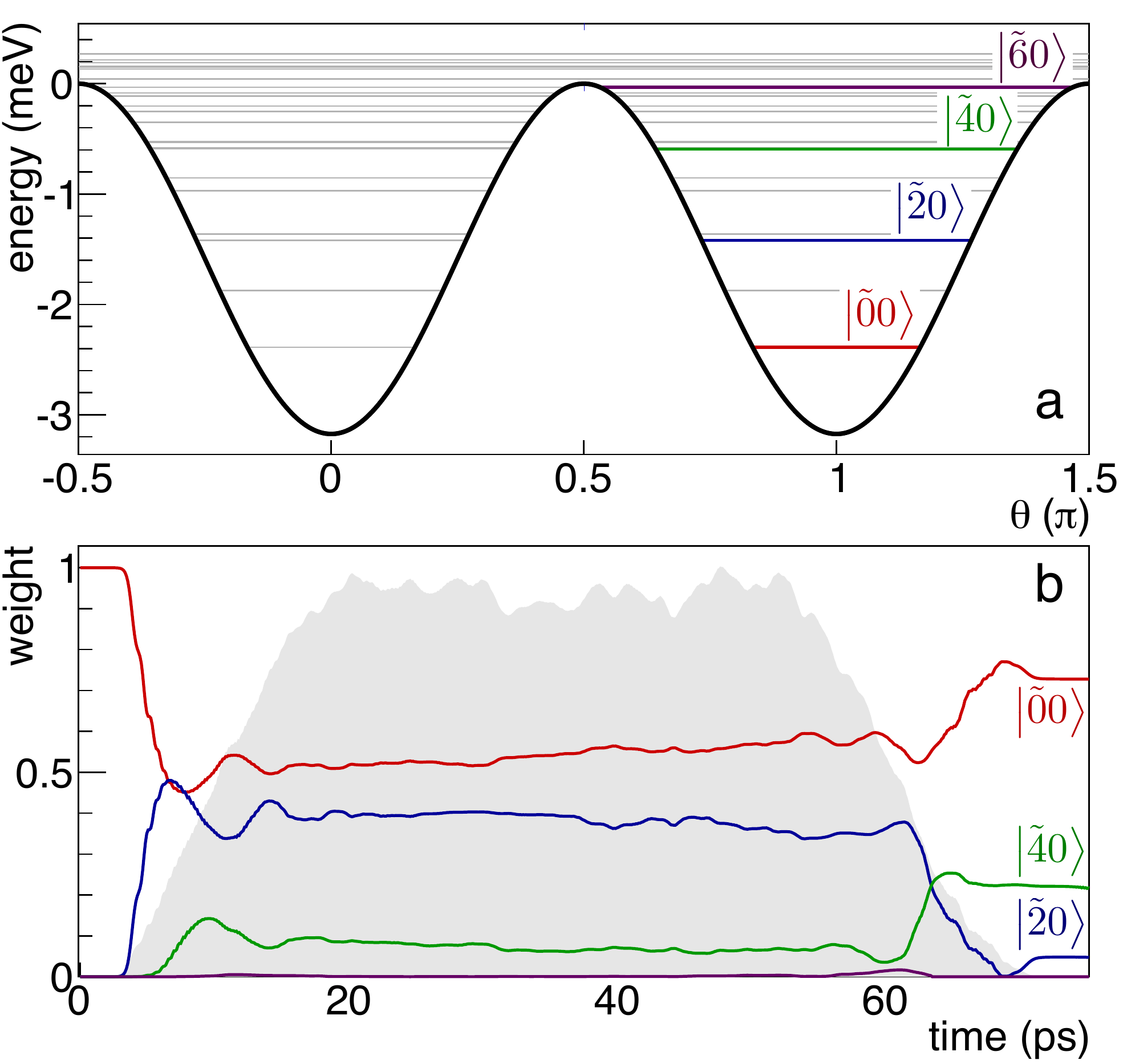}
   \caption{a) Potential energy curve (black line) and energies of the pendular states in the
      depicted energy range (grey and colored lines) for $I=6\cdot10^{11}$~W/cm$^2$. The
      experimentally populated adiabatic pendular states are shown in color and are labeled by
      \ket{\tilde{J},M}. b) The time-dependent weights of the projections of the pendular wave
      packet, which started from the field-free ground state, onto the pendular-state basis. The
      temporal laser profile is indicated by the gray area.}
   \label{fig:level-time}
\end{figure}

In our theoretical description, we solved the time-dependent Schrödinger
equation~\cite{SanchezMoreno:PRA76:053413} starting in the field-free rotational ground state and
using the experimental temporal profile of the laser intensity. The angular part of the interaction
potential between the molecules and the nonresonant laser field,
$-I(t)\,\Delta\alpha\cos^2\!\theta/(2c\epsilon_0)$, is presented in \autoref{fig:level-time}\,a for
$I=6\cdot10^{11}\text{~W/cm}^2$; $I$ is the laser intensity, $\Delta\alpha$ the polarizability
anisotropy, and $\theta$ the angle between the alignment laser polarization, $Y$, and the axis of
the molecule, $z$. Moreover, the energies of the pendular states \ket{\tilde{J},M} in that potential
are depicted. Here, the \ket{\tilde{J},M} labels are used for the pendular states that correlate
adiabatically with the field free rotational states \ket{J,M}.
To rationalize the experimental observations, we computed the degree of alignment $\cost$ and the
decomposition of the wave packet in terms of the adiabatic pendular eigenstates. Our theoretical
description includes the velocity distribution of the ions after the Coulomb explosion and a volume
effect model which takes into account the spatial intensity profiles of the alignment and the probe
laser pulses~\cite{Omiste:PCCP13:18815}.

\begin{figure}
   \centering
   \includegraphics[width=\figwidth]{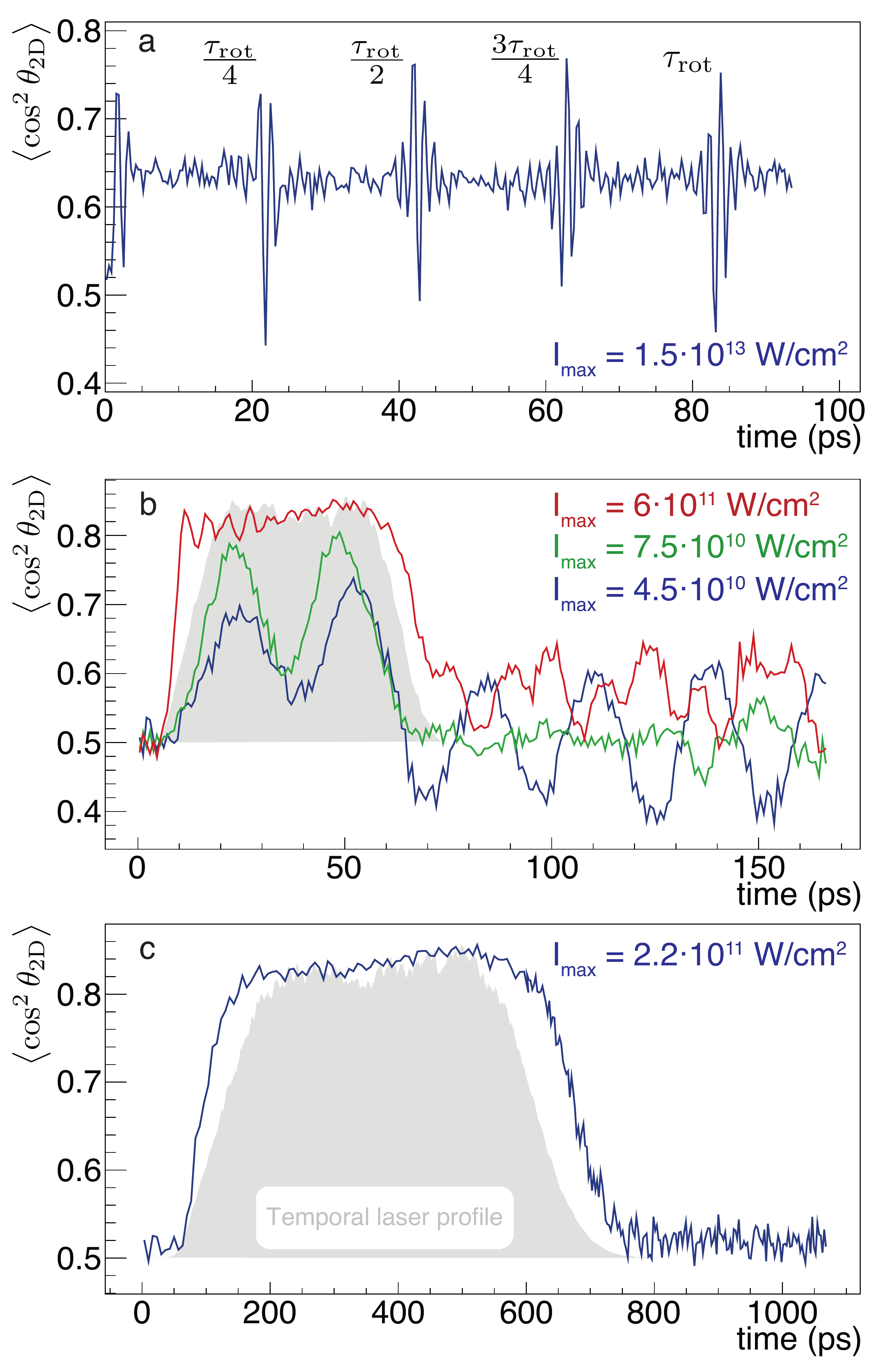}
   \caption{Measured degree of alignment \cost as a function of the delay between the alignment and
      probe laser pulses for an alignment pulse duration of a)~450~fs, b)~50~ps and c)~485~ps. The
      temporal profiles of the alignment laser pulses are indicated by the gray areas.}
   \label{fig:LineProfiles}
\end{figure}

\autoref{fig:LineProfiles}\,a shows the degree of alignment measured as a function of the delay
between the alignment and probe laser pulses for an alignment pulse duration of 450~fs (impulsive
alignment) and a peak intensity of $1.5\cdot10^{13}$~W/cm$^2$.
These results fully agree with previous experiments and the analysis of the prominent quarter-period
revival confirms that the molecules are prepared in the absolute ground state
$J=0$~\cite{Nielsen:PCCP13:18971}.

\autoref{fig:LineProfiles}\,c shows the degree of alignment for a pulse duration of 485~ps and a
peak intensity of $2.2\cdot10^{11}$~W/cm$^2$. The temporal laser profile is indicated by the shaded
area. The rise- and fall-times of the pulse are 100~ps and 150~ps, respectively (10\,\%--90\,\%).
The degree of alignment adiabatically follows the temporal laser profile as expected for pulses
where all relevant time scales are larger than the rotational period of the molecule.

\autoref{fig:LineProfiles}\,b shows the time dependence of the degree of alignment for a pulse
duration of 50~ps (FWHM in intensity) and intensities \mbox{$4.5\cdot10^{10}$~W/cm$^2$} (blue),
\mbox{$7.5\cdot10^{10}$~W/cm$^2$} (green) and \mbox{$6\cdot10^{11}$~W/cm$^2$} (red). The
10\,\%--90\,\% rise time of the laser pulse is 10~ps. It is followed by a plateau where the laser
intensity is approximately constant. With $\taurot\approx82$~ps this places the temporal features of
the pulse between adiabatic and impulsive alignment. The rising edge of the laser pulse creates a
wave packet of pendular states through Raman-coupling with selection rules $\Delta{J}=0,\pm2$ and
$\Delta{}M=0$. This wave packet propagates in the effective potential for the molecules in the laser
field, giving rise to an oscillatory modulation of the degree of alignment. These oscillations are
attributed to the bouncing back and forth of the wave packet in the potential. A movie depicting the
wave-packet dynamics is provided in the supplementary information. They resemble the librational
motion of a classical pendulum.

The oscillation frequency increases with increasing peak intensity of the alignment laser,
indicating the admixing of higher-angular-momentum states. Simultaneously, the amplitude of the
oscillation decreases, depicting the stronger angular confinement of the molecules deeper in the
potential and, therefore, a smaller change in the degree of alignment. The oscillations are very
pronounced at the beginning of the laser pulse, but their amplitude decreases toward the end of the
pulse. Initially, the phase of oscillation is defined by the rising edge, \ie, it is nearly the same
for all molecules. The decrease during the laser pulse is mainly attributed to the volume effect,
\ie, different molecules experiencing different laser intensities~\cite{Omiste:PCCP13:18815}.
Additional contributions to the decrease of contrast in the amplitude of the oscillations are due to
the not completely flat temporal laser-intensity profile and the anharmonic-oscillator potential,
\autoref{fig:level-time}\,a.

\begin{figure}[t]
   \centering
   \includegraphics[width=\linewidth]{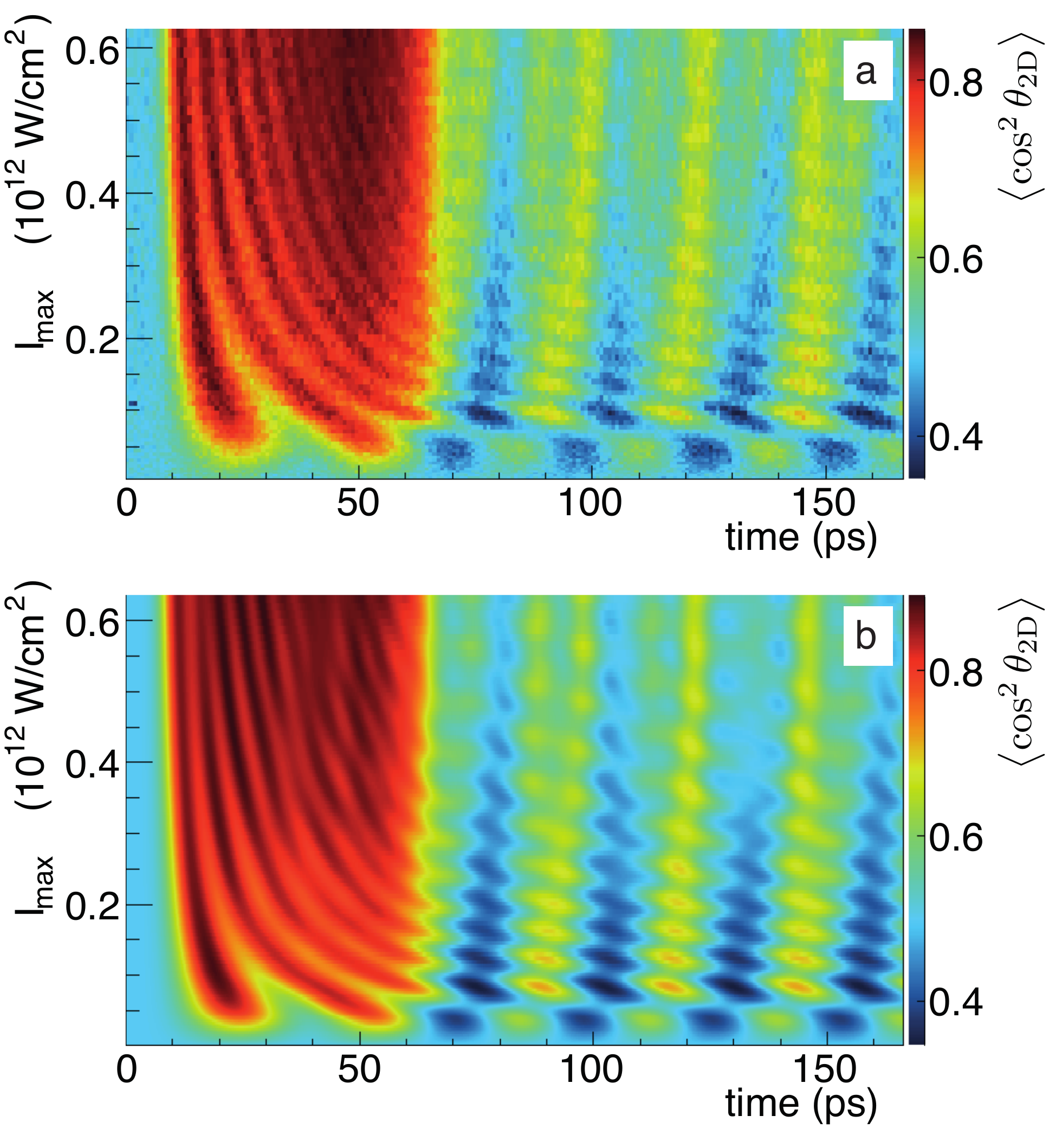}
   \caption{\cost as a function of the delay between the alignment and the probe laser pulses and
      the peak intensity of the alignment laser pulse. The alignment pulse duration is 50~ps. a)
      Experimental, b) theoretical results.}
   \label{fig:50ps_2D}
\end{figure}

The calculated decomposition of the wave packet in terms of its pendular-state basis for the 50~ps
pulse of intensity \mbox{$6\cdot10^{11}$~W/cm$^2$}, shown in \autoref{fig:LineProfiles}\,b, is given
in \autoref{fig:level-time}\,b. The coefficients show rapid changes at the two edges of the pulse,
whereas they keep an approximately constant value in the plateau region. Here, only
\ket{\tilde{0},0}, \ket{\tilde{2},0}, and \ket{\tilde{4},0} contribute significantly to the
dynamics. The oscillations in the degree of alignment are due to the temporal evolution of the phase
of these pendular states. \autoref{fig:level-time}\,a illustrates that during the pulse all
contributing states are bound in the potential well. For the $4.5\cdot10^{10}$~W/cm$^2$ and
$7.5\cdot10^{10}$~W/cm$^2$ pulses the pendular ground state has by far the largest contribution
($\gtrsim0.89$) to the wave packet. Thus, the oscillations in \cost are due to the coherence between
\ket{\tilde{0},0} and \ket{\tilde{2},0}, with the latter being unbound for these intensities. These
two-state interferences are reflected by the cosine-like oscillations of the degree of alignment.

To obtain further insight into the alignment dynamics in the intermediate regime the degree of
alignment is recorded for a range of alignment pulse peak intensities as a function of time. The
alignment pulse duration is 50~ps as in \autoref{fig:LineProfiles}\,b. A 2D representation of the
experimental results is shown in \autoref{fig:50ps_2D}\,a. The corresponding theoretical
calculations are shown in \autoref{fig:50ps_2D}\,b. The oscillatory behavior of the degree of
alignment during the laser pulse (between 10~ps and 50~ps) is again strongly visible. The
frequencies of the oscillations are small in the low intensity regime (\emph{vide supra}). As the
laser intensity is increased more and more oscillations are observed.

\autoref{fig:50ps_2D} also shows a complex behavior of the degree of alignment as soon as the laser
is switched off, with a strong dependence on the intensity of the laser pulse. At low laser
intensities we observed a revival structure corresponding to a single-cosine dependence due to the
beating of $\ket{0,0}$ and $\ket{2,0}$. Moreover, for an intensity of $7.5\cdot10^{10}$~W/cm$^2$, we
find that the revival structure is strongly suppressed and only a weak revival structure is
observed; this is also visible in the green trace in \autoref{fig:LineProfiles}\,b. For these
conditions, the field-free state after the laser pulse closely resembled the initial field-free
state~\cite{Ortigoso:JCP110:3870}. The phase between wave packet components is modified by the
falling edge of the laser pulse in such a way that the revival structure is coherently switched off.
This effect repeats itself for increasing laser intensities. These quantum interferences are similar
to those previously observed with two appropriately delayed laser pulses~\cite{Lee:PRL93:233601},
but now it is achieved with a single pulse of appropriate duration and intensity. The computational
results in \autoref{fig:50ps_2D}\,b are in excellent agreement with the experimental results in
\autoref{fig:50ps_2D}\,a. For \mbox{$I_\text{max}\approx6\cdot10^{10}$~W/cm$^2$} these computations
predict no alignment once the pulse is off with small oscillations around the mean value of
$\cost\approx0.50$. The incomplete suppression of the experimental alignment structure is attributed
to the volume effect and, thus, the simultaneous observation of dynamics for different
field-strengths.

In conclusion, we studied the time dependent alignment behavior of state-selected OCS molecules in
their absolute ground state for a pulse duration in the intermediate regime, between impulsive and
adiabatic alignment. We observed strong oscillations in the degree of alignment during the laser
pulse. These oscillations are attributed to the propagation of a wave packet in the potential of the
molecule in the alignment laser field. The observed motion is the quantum analogue of an
oscillating pendulum. Our results show the opportunity to initiate and to stop wave packet motion
within a short period and by a single laser pulse. It provides an effective coherent control scheme
of molecular motion.

The wave packet dynamics inside the effective potential has implications on the performance of
experiments with laser aligned molecules such as the investigation of molecular-frame photoelectron
angular distributions~\cite{Holmegaard:NatPhys6:428}, the detection of structural changes via X-ray
and electron diffraction~\cite{Spence:PRL92:198102, Filsinger:PCCP13:2076, Hensley:PRL109:133202,
   Kuepper:LCLSdibn:inprep} and photoelectron holography from within~\cite{Boll:PRA88:061402},
because typically a strong degree of alignment is required for these experiments. Reducing the
jitter of the relative timing and applying the appropriate delay of the alignment and probe laser
allows for the strongest degree of alignment to be achieved, since the molecules can be probed at
one of the maxima of \costhreeD of the pendular state dynamics. This holds especially for large
molecules with rotational periods on the order of a few nanoseconds (ns), where non adiabatic
effects will start to play a role even with ns laser pulses that are often employed to strongly fix
molecules in space.

The observed pendular motion has implications on the performance of ultrafast molecular switches
based on internal-rotation dynamics~\cite{Choi:PRL96:156106, Avellini:ACIE51:1611}. The torsional
motion of non-rigid quantum objects~\cite{Ramakrishna:PRL99:103001, Madsen:PRL102:073007} or surface
adsorbed molecules~\cite{Choi:PRL96:156106, Donhauser:Science292:2303, Reuter:PRL101:208303} is
governed by a $2\pi$-periodic potential about the torsional or dihedral
angle~\cite{Gordy:MWMolSpec}. Variations in the relative alignment of the two moieties will lead to
variations in, \eg, the current through a molecular switch~\cite{delValle:NatNano2:176,
   Ramakrishna:PRL99:103001}. Inducing a wave packet of the internal rotation to coherently switch
the system and terminating the motion in the desired position by the end of the laser
pulse~\cite{Grossmann:EPL60:201, Barbatti:CP350:145, Parker:JCP135:224301} would work even when the
surrounding media is not dissipative.

This work has been supported by the excellence cluster ``The Hamburg Center for Ultrafast Imaging --
Structure, Dynamics and Control of Matter at the Atomic Scale'' of the Deutsche
Forschungsgemeinschaft, including the Mildred Dresselhaus award for R.G.F. ~ R.G.F. also gratefully
acknowledges financial support by the Spanish Ministry of Science FIS2011-24540 (MICINN), the grants
P11-FQM-7276 and FQM-4643 (Junta de Andaluc\'{\i}a), and by the Andalusian research group FQM-207.
J.J.O.\ acknowledges support through the research plan of the Universidad de Granada. N.L.M.M.\
gratefully acknowledges a fellowship of the Joachim Herz Stiftung.

\bibliography{string,cmi}

\begin{thebibliography}{37}%
\makeatletter
\providecommand \@ifxundefined [1]{%
 \@ifx{#1\undefined}
}%
\providecommand \@ifnum [1]{%
 \ifnum #1\expandafter \@firstoftwo
 \else \expandafter \@secondoftwo
 \fi
}%
\providecommand \@ifx [1]{%
 \ifx #1\expandafter \@firstoftwo
 \else \expandafter \@secondoftwo
 \fi
}%
\providecommand \natexlab [1]{#1}%
\providecommand \enquote  [1]{``#1''}%
\providecommand \bibnamefont  [1]{#1}%
\providecommand \bibfnamefont [1]{#1}%
\providecommand \citenamefont [1]{#1}%
\providecommand \href@noop [0]{\@secondoftwo}%
\providecommand \href [0]{\begingroup \@sanitize@url \@href}%
\providecommand \@href[1]{\@@startlink{#1}\@@href}%
\providecommand \@@href[1]{\endgroup#1\@@endlink}%
\providecommand \@sanitize@url [0]{\catcode `\\12\catcode `\$12\catcode
  `\&12\catcode `\#12\catcode `\^12\catcode `\_12\catcode `\%12\relax}%
\providecommand \@@startlink[1]{}%
\providecommand \@@endlink[0]{}%
\providecommand \url  [0]{\begingroup\@sanitize@url \@url }%
\providecommand \@url [1]{\endgroup\@href {#1}{\urlprefix }}%
\providecommand \urlprefix  [0]{URL }%
\providecommand \Eprint [0]{\href }%
\providecommand \doibase [0]{http://dx.doi.org/}%
\providecommand \selectlanguage [0]{\@gobble}%
\providecommand \bibinfo  [0]{\@secondoftwo}%
\providecommand \bibfield  [0]{\@secondoftwo}%
\providecommand \translation [1]{[#1]}%
\providecommand \BibitemOpen [0]{}%
\providecommand \bibitemStop [0]{}%
\providecommand \bibitemNoStop [0]{.\EOS\space}%
\providecommand \EOS [0]{\spacefactor3000\relax}%
\providecommand \BibitemShut  [1]{\csname bibitem#1\endcsname}%
\let\auto@bib@innerbib\@empty
\bibitem [{\citenamefont {Loesch}\ and\ \citenamefont
  {Remscheid}(1990)}]{Loesch:JCP93:4779}%
  \BibitemOpen
  \bibfield  {author} {\bibinfo {author} {\bibfnamefont {H.~J.}\ \bibnamefont
  {Loesch}}\ and\ \bibinfo {author} {\bibfnamefont {A.}~\bibnamefont
  {Remscheid}},\ }\bibfield  {title} {\enquote {\bibinfo {title} {Brute force
  in molecular reaction dynamics: {A} novel technique for measuring steric
  effects},}\ }\href {\doibase 10.1063/1.458668} {\bibfield  {journal}
  {\bibinfo  {journal} {J.\ Chem.\ Phys.}\ }\textbf {\bibinfo {volume} {93}},\
  \bibinfo {pages} {4779} (\bibinfo {year} {1990})}\BibitemShut {NoStop}%
\bibitem [{\citenamefont {Friedrich}\ and\ \citenamefont
  {Herschbach}(1991)}]{Friedrich:Nature353:412}%
  \BibitemOpen
  \bibfield  {author} {\bibinfo {author} {\bibfnamefont {B.}~\bibnamefont
  {Friedrich}}\ and\ \bibinfo {author} {\bibfnamefont {D.~R.}\ \bibnamefont
  {Herschbach}},\ }\bibfield  {title} {\enquote {\bibinfo {title} {Spatial
  orientation of molecules in strong electric fields and evidence for pendular
  states},}\ }\href {\doibase 10.1038/353412a0} {\bibfield  {journal} {\bibinfo
   {journal} {Nature}\ }\textbf {\bibinfo {volume} {353}},\ \bibinfo {pages}
  {412--414} (\bibinfo {year} {1991})}\BibitemShut {NoStop}%
\bibitem [{\citenamefont {Friedrich}\ and\ \citenamefont
  {Herschbach}(1995)}]{Friedrich:PRL74:4623}%
  \BibitemOpen
  \bibfield  {author} {\bibinfo {author} {\bibfnamefont {Bretislav}\
  \bibnamefont {Friedrich}}\ and\ \bibinfo {author} {\bibfnamefont {Dudley}\
  \bibnamefont {Herschbach}},\ }\bibfield  {title} {\enquote {\bibinfo {title}
  {Alignment and trapping of molecules in intense laser fields},}\ }\href
  {\doibase 10.1103/PhysRevLett.74.4623} {\bibfield  {journal} {\bibinfo
  {journal} {Phys.\ Rev.\ Lett.}\ }\textbf {\bibinfo {volume} {74}},\ \bibinfo
  {pages} {4623--4626} (\bibinfo {year} {1995})}\BibitemShut {NoStop}%
\bibitem [{\citenamefont {Stapelfeldt}\ and\ \citenamefont
  {Seideman}(2003)}]{Stapelfeldt:RMP75:543}%
  \BibitemOpen
  \bibfield  {author} {\bibinfo {author} {\bibfnamefont {Henrik}\ \bibnamefont
  {Stapelfeldt}}\ and\ \bibinfo {author} {\bibfnamefont {Tamar}\ \bibnamefont
  {Seideman}},\ }\bibfield  {title} {\enquote {\bibinfo {title} {Colloquium:
  Aligning molecules with strong laser pulses},}\ }\href {\doibase
  10.1103/RevModPhys.75.543} {\bibfield  {journal} {\bibinfo  {journal} {Rev.\
  Mod.\ Phys.}\ }\textbf {\bibinfo {volume} {75}},\ \bibinfo {pages} {543--557}
  (\bibinfo {year} {2003})}\BibitemShut {NoStop}%
\bibitem [{\citenamefont {Nauta}\ and\ \citenamefont
  {Miller}(1999)}]{Nauta:Science283:1895}%
  \BibitemOpen
  \bibfield  {author} {\bibinfo {author} {\bibfnamefont {Klaas}\ \bibnamefont
  {Nauta}}\ and\ \bibinfo {author} {\bibfnamefont {Roger~E.}\ \bibnamefont
  {Miller}},\ }\bibfield  {title} {\enquote {\bibinfo {title} {Nonequilibrium
  self-assembly of long chains of polar molecules in superfluid helium},}\
  }\href {\doibase 10.1126/science.283.5409.1895} {\bibfield  {journal}
  {\bibinfo  {journal} {Science}\ }\textbf {\bibinfo {volume} {283}},\ \bibinfo
  {pages} {1895} (\bibinfo {year} {1999})}\BibitemShut {NoStop}%
\bibitem [{\citenamefont {Spence}\ and\ \citenamefont
  {Doak}(2004)}]{Spence:PRL92:198102}%
  \BibitemOpen
  \bibfield  {author} {\bibinfo {author} {\bibfnamefont {John C~H}\
  \bibnamefont {Spence}}\ and\ \bibinfo {author} {\bibfnamefont {R~B}\
  \bibnamefont {Doak}},\ }\bibfield  {title} {\enquote {\bibinfo {title}
  {Single molecule diffraction},}\ }\href {\doibase
  10.1103/PhysRevLett.92.198102} {\bibfield  {journal} {\bibinfo  {journal}
  {Phys.\ Rev.\ Lett.}\ }\textbf {\bibinfo {volume} {92}},\ \bibinfo {pages}
  {198102} (\bibinfo {year} {2004})}\BibitemShut {NoStop}%
\bibitem [{\citenamefont {Madsen}\ \emph {et~al.}(2009)\citenamefont {Madsen},
  \citenamefont {Madsen}, \citenamefont {Viftrup}, \citenamefont {Johansson},
  \citenamefont {Poulsen}, \citenamefont {Holmegaard}, \citenamefont
  {Kumarappan}, \citenamefont {J{\o}rgensen},\ and\ \citenamefont
  {Stapelfeldt}}]{Madsen:PRL102:073007}%
  \BibitemOpen
  \bibfield  {author} {\bibinfo {author} {\bibfnamefont {C.~B.}\ \bibnamefont
  {Madsen}}, \bibinfo {author} {\bibfnamefont {L.~B.}\ \bibnamefont {Madsen}},
  \bibinfo {author} {\bibfnamefont {S.~S.}\ \bibnamefont {Viftrup}}, \bibinfo
  {author} {\bibfnamefont {M.~P.}\ \bibnamefont {Johansson}}, \bibinfo {author}
  {\bibfnamefont {T.~B.}\ \bibnamefont {Poulsen}}, \bibinfo {author}
  {\bibfnamefont {L.}~\bibnamefont {Holmegaard}}, \bibinfo {author}
  {\bibfnamefont {V.}~\bibnamefont {Kumarappan}}, \bibinfo {author}
  {\bibfnamefont {K.A.}\ \bibnamefont {J{\o}rgensen}}, \ and\ \bibinfo {author}
  {\bibfnamefont {H.}~\bibnamefont {Stapelfeldt}},\ }\bibfield  {title}
  {\enquote {\bibinfo {title} {Manipulating the torsion of molecules by strong
  laser pulses},}\ }\href {\doibase 10.1103/PhysRevLett.102.073007} {\bibfield
  {journal} {\bibinfo  {journal} {Phys.\ Rev.\ Lett.}\ }\textbf {\bibinfo
  {volume} {102}},\ \bibinfo {pages} {073007} (\bibinfo {year} {2009})},\
  \Eprint {http://arxiv.org/abs/0809.2935} {arXiv:0809.2935 [physics]}
  \BibitemShut {NoStop}%
\bibitem [{\citenamefont {Purcell}\ and\ \citenamefont
  {Barker}(2009)}]{Purcell:PRL103:153001}%
  \BibitemOpen
  \bibfield  {author} {\bibinfo {author} {\bibfnamefont {S~M}\ \bibnamefont
  {Purcell}}\ and\ \bibinfo {author} {\bibfnamefont {P~F}\ \bibnamefont
  {Barker}},\ }\bibfield  {title} {\enquote {\bibinfo {title} {{Tailoring the
  Optical Dipole Force for Molecules by Field-Induced Alignment}},}\ }\href
  {\doibase 10.1103/PhysRevLett.103.153001} {\bibfield  {journal} {\bibinfo
  {journal} {Phys. Rev. Lett.}\ }\textbf {\bibinfo {volume} {103}},\ \bibinfo
  {pages} {153001} (\bibinfo {year} {2009})}\BibitemShut {NoStop}%
\bibitem [{\citenamefont {Holmegaard}\ \emph {et~al.}(2010)\citenamefont
  {Holmegaard}, \citenamefont {Hansen}, \citenamefont {Kalh{\o}j},
  \citenamefont {Kragh}, \citenamefont {Stapelfeldt}, \citenamefont
  {Filsinger}, \citenamefont {K\"upper}, \citenamefont {Meijer}, \citenamefont
  {Dimitrovski}, \citenamefont {Abu-samha}, \citenamefont {Martiny},\ and\
  \citenamefont {Madsen}}]{Holmegaard:NatPhys6:428}%
  \BibitemOpen
  \bibfield  {author} {\bibinfo {author} {\bibfnamefont {Lotte}\ \bibnamefont
  {Holmegaard}}, \bibinfo {author} {\bibfnamefont {Jonas~L.}\ \bibnamefont
  {Hansen}}, \bibinfo {author} {\bibfnamefont {Line}\ \bibnamefont
  {Kalh{\o}j}}, \bibinfo {author} {\bibfnamefont {Sofie~Louise}\ \bibnamefont
  {Kragh}}, \bibinfo {author} {\bibfnamefont {Henrik}\ \bibnamefont
  {Stapelfeldt}}, \bibinfo {author} {\bibfnamefont {Frank}\ \bibnamefont
  {Filsinger}}, \bibinfo {author} {\bibfnamefont {Jochen}\ \bibnamefont
  {K\"upper}}, \bibinfo {author} {\bibfnamefont {Gerard}\ \bibnamefont
  {Meijer}}, \bibinfo {author} {\bibfnamefont {Darko}\ \bibnamefont
  {Dimitrovski}}, \bibinfo {author} {\bibfnamefont {Mahmoud}\ \bibnamefont
  {Abu-samha}}, \bibinfo {author} {\bibfnamefont {Christian P.~J.}\
  \bibnamefont {Martiny}}, \ and\ \bibinfo {author} {\bibfnamefont
  {Lars~Bojer}\ \bibnamefont {Madsen}},\ }\bibfield  {title} {\enquote
  {\bibinfo {title} {Photoelectron angular distributions from strong-field
  ionization of oriented molecules},}\ }\href {\doibase 10.1038/NPHYS1666}
  {\bibfield  {journal} {\bibinfo  {journal} {Nat. Phys.}\ }\textbf {\bibinfo
  {volume} {6}},\ \bibinfo {pages} {428} (\bibinfo {year} {2010})},\ \Eprint
  {http://arxiv.org/abs/1003.4634} {arXiv:1003.4634 [physics]} \BibitemShut
  {NoStop}%
\bibitem [{\citenamefont {Boll}\ \emph {et~al.}(2013)\citenamefont {Boll},
  \citenamefont {Anielski}, \citenamefont {Bostedt}, \citenamefont {Bozek},
  \citenamefont {Christensen}, \citenamefont {Coffee}, \citenamefont {De},
  \citenamefont {Decleva}, \citenamefont {Epp}, \citenamefont {Erk},
  \citenamefont {Foucar}, \citenamefont {Krasniqi}, \citenamefont {K\"upper},
  \citenamefont {Rouz\'{e}e}, \citenamefont {Rudek}, \citenamefont {Rudenko},
  \citenamefont {Schorb}, \citenamefont {Stapelfeldt}, \citenamefont {Stener},
  \citenamefont {Stern}, \citenamefont {Techert}, \citenamefont {Trippel},
  \citenamefont {Vrakking}, \citenamefont {Ullrich},\ and\ \citenamefont
  {Rolles}}]{Boll:PRA88:061402}%
  \BibitemOpen
  \bibfield  {author} {\bibinfo {author} {\bibfnamefont {Rebecca}\ \bibnamefont
  {Boll}}, \bibinfo {author} {\bibfnamefont {Dennis}\ \bibnamefont {Anielski}},
  \bibinfo {author} {\bibfnamefont {Christoph}\ \bibnamefont {Bostedt}},
  \bibinfo {author} {\bibfnamefont {John~D.}\ \bibnamefont {Bozek}}, \bibinfo
  {author} {\bibfnamefont {Lauge}\ \bibnamefont {Christensen}}, \bibinfo
  {author} {\bibfnamefont {Ryan}\ \bibnamefont {Coffee}}, \bibinfo {author}
  {\bibfnamefont {Sankar}\ \bibnamefont {De}}, \bibinfo {author} {\bibfnamefont
  {Piero}\ \bibnamefont {Decleva}}, \bibinfo {author} {\bibfnamefont
  {Sascha~W.}\ \bibnamefont {Epp}}, \bibinfo {author} {\bibfnamefont
  {Benjamin}\ \bibnamefont {Erk}}, \bibinfo {author} {\bibfnamefont {Lutz}\
  \bibnamefont {Foucar}}, \bibinfo {author} {\bibfnamefont {Faton}\
  \bibnamefont {Krasniqi}}, \bibinfo {author} {\bibfnamefont {Jochen}\
  \bibnamefont {K\"upper}}, \bibinfo {author} {\bibfnamefont {Arnaud}\
  \bibnamefont {Rouz\'{e}e}}, \bibinfo {author} {\bibfnamefont {Benedikt}\
  \bibnamefont {Rudek}}, \bibinfo {author} {\bibfnamefont {Artem}\ \bibnamefont
  {Rudenko}}, \bibinfo {author} {\bibfnamefont {Sebastian}\ \bibnamefont
  {Schorb}}, \bibinfo {author} {\bibfnamefont {Henrik}\ \bibnamefont
  {Stapelfeldt}}, \bibinfo {author} {\bibfnamefont {Mauro}\ \bibnamefont
  {Stener}}, \bibinfo {author} {\bibfnamefont {Stephan}\ \bibnamefont {Stern}},
  \bibinfo {author} {\bibfnamefont {Simone}\ \bibnamefont {Techert}}, \bibinfo
  {author} {\bibfnamefont {Sebastian}\ \bibnamefont {Trippel}}, \bibinfo
  {author} {\bibfnamefont {Marc J.~J.}\ \bibnamefont {Vrakking}}, \bibinfo
  {author} {\bibfnamefont {Joachim}\ \bibnamefont {Ullrich}}, \ and\ \bibinfo
  {author} {\bibfnamefont {Daniel}\ \bibnamefont {Rolles}},\ }\bibfield
  {title} {\enquote {\bibinfo {title} {Femtosecond photoelectron diffraction on
  laser-aligned molecules: Towards time-resolved imaging of molecular
  structure},}\ }\href {\doibase 10.1103/PhysRevA.88.061402} {\bibfield
  {journal} {\bibinfo  {journal} {Phys.\ Rev.\ A}\ }\textbf {\bibinfo {volume}
  {88}},\ \bibinfo {pages} {061402} (\bibinfo {year} {2013})}\BibitemShut
  {NoStop}%
\bibitem [{\citenamefont {K{\"u}pper}\ \emph {et~al.}(2013)\citenamefont
  {K{\"u}pper}, \citenamefont {Stern}, \citenamefont {Holmegaard},
  \citenamefont {Filsinger}, \citenamefont {Rouz\'{e}e}, \citenamefont
  {Rudenko}, \citenamefont {Johnsson}, \citenamefont {Martin}, \citenamefont
  {Adolph}, \citenamefont {Aquila}, \citenamefont {Bajt}, \citenamefont
  {Barty}, \citenamefont {Bostedt}, \citenamefont {Bozek}, \citenamefont
  {Caleman}, \citenamefont {Coffee}, \citenamefont {Coppola}, \citenamefont
  {Delmas}, \citenamefont {Epp}, \citenamefont {Erk}, \citenamefont {Foucar},
  \citenamefont {Gorkhover}, \citenamefont {Gumprecht}, \citenamefont
  {Hartmann}, \citenamefont {Hartmann}, \citenamefont {Hauser}, \citenamefont
  {Holl}, \citenamefont {H{\"o}mke}, \citenamefont {Kimmel}, \citenamefont
  {Krasniqi}, \citenamefont {K{\"u}hnel}, \citenamefont {Maurer}, \citenamefont
  {Messerschmidt}, \citenamefont {Moshammer}, \citenamefont {Reich},
  \citenamefont {Rudek}, \citenamefont {Santra}, \citenamefont {Schlichting},
  \citenamefont {Schmidt}, \citenamefont {Schorb}, \citenamefont {Schulz},
  \citenamefont {Soltau}, \citenamefont {Spence}, \citenamefont {Starodub},
  \citenamefont {Str{\"u}der}, \citenamefont {Th{\o}gersen}, \citenamefont
  {Vrakking}, \citenamefont {Weidenspointner}, \citenamefont {White},
  \citenamefont {Wunderer}, \citenamefont {Meijer}, \citenamefont {Ullrich},
  \citenamefont {Stapelfeldt}, \citenamefont {Rolles},\ and\ \citenamefont
  {Chapman}}]{Kuepper:LCLSdibn:inprep}%
  \BibitemOpen
  \bibfield  {author} {\bibinfo {author} {\bibfnamefont {Jochen}\ \bibnamefont
  {K{\"u}pper}}, \bibinfo {author} {\bibfnamefont {Stephan}\ \bibnamefont
  {Stern}}, \bibinfo {author} {\bibfnamefont {Lotte}\ \bibnamefont
  {Holmegaard}}, \bibinfo {author} {\bibfnamefont {Frank}\ \bibnamefont
  {Filsinger}}, \bibinfo {author} {\bibfnamefont {Arnaud}\ \bibnamefont
  {Rouz\'{e}e}}, \bibinfo {author} {\bibfnamefont {Artem}\ \bibnamefont
  {Rudenko}}, \bibinfo {author} {\bibfnamefont {Per}\ \bibnamefont {Johnsson}},
  \bibinfo {author} {\bibfnamefont {Andrew~V.}\ \bibnamefont {Martin}},
  \bibinfo {author} {\bibfnamefont {Marcus}\ \bibnamefont {Adolph}}, \bibinfo
  {author} {\bibfnamefont {Andrew}\ \bibnamefont {Aquila}}, \bibinfo {author}
  {\bibfnamefont {Sa{\v s}a}\ \bibnamefont {Bajt}}, \bibinfo {author}
  {\bibfnamefont {Anton}\ \bibnamefont {Barty}}, \bibinfo {author}
  {\bibfnamefont {Christoph}\ \bibnamefont {Bostedt}}, \bibinfo {author}
  {\bibfnamefont {John}\ \bibnamefont {Bozek}}, \bibinfo {author}
  {\bibfnamefont {Carl}\ \bibnamefont {Caleman}}, \bibinfo {author}
  {\bibfnamefont {Ryan}\ \bibnamefont {Coffee}}, \bibinfo {author}
  {\bibfnamefont {Nicola}\ \bibnamefont {Coppola}}, \bibinfo {author}
  {\bibfnamefont {Tjark}\ \bibnamefont {Delmas}}, \bibinfo {author}
  {\bibfnamefont {Sascha}\ \bibnamefont {Epp}}, \bibinfo {author}
  {\bibfnamefont {Benjamin}\ \bibnamefont {Erk}}, \bibinfo {author}
  {\bibfnamefont {Lutz}\ \bibnamefont {Foucar}}, \bibinfo {author}
  {\bibfnamefont {Tais}\ \bibnamefont {Gorkhover}}, \bibinfo {author}
  {\bibfnamefont {Lars}\ \bibnamefont {Gumprecht}}, \bibinfo {author}
  {\bibfnamefont {Andreas}\ \bibnamefont {Hartmann}}, \bibinfo {author}
  {\bibfnamefont {Robert}\ \bibnamefont {Hartmann}}, \bibinfo {author}
  {\bibfnamefont {G{\"u}nter}\ \bibnamefont {Hauser}}, \bibinfo {author}
  {\bibfnamefont {Peter}\ \bibnamefont {Holl}}, \bibinfo {author}
  {\bibfnamefont {Andre}\ \bibnamefont {H{\"o}mke}}, \bibinfo {author}
  {\bibfnamefont {Nils}\ \bibnamefont {Kimmel}}, \bibinfo {author}
  {\bibfnamefont {Faton}\ \bibnamefont {Krasniqi}}, \bibinfo {author}
  {\bibfnamefont {Kai-Uwe}\ \bibnamefont {K{\"u}hnel}}, \bibinfo {author}
  {\bibfnamefont {Jochen}\ \bibnamefont {Maurer}}, \bibinfo {author}
  {\bibfnamefont {Marc}\ \bibnamefont {Messerschmidt}}, \bibinfo {author}
  {\bibfnamefont {Robert}\ \bibnamefont {Moshammer}}, \bibinfo {author}
  {\bibfnamefont {Christian}\ \bibnamefont {Reich}}, \bibinfo {author}
  {\bibfnamefont {Benedikt}\ \bibnamefont {Rudek}}, \bibinfo {author}
  {\bibfnamefont {Robin}\ \bibnamefont {Santra}}, \bibinfo {author}
  {\bibfnamefont {Ilme}\ \bibnamefont {Schlichting}}, \bibinfo {author}
  {\bibfnamefont {Carlo}\ \bibnamefont {Schmidt}}, \bibinfo {author}
  {\bibfnamefont {Sebastian}\ \bibnamefont {Schorb}}, \bibinfo {author}
  {\bibfnamefont {Joachim}\ \bibnamefont {Schulz}}, \bibinfo {author}
  {\bibfnamefont {Heike}\ \bibnamefont {Soltau}}, \bibinfo {author}
  {\bibfnamefont {John C.~H.}\ \bibnamefont {Spence}}, \bibinfo {author}
  {\bibfnamefont {Dmitri}\ \bibnamefont {Starodub}}, \bibinfo {author}
  {\bibfnamefont {Lothar}\ \bibnamefont {Str{\"u}der}}, \bibinfo {author}
  {\bibfnamefont {Jan}\ \bibnamefont {Th{\o}gersen}}, \bibinfo {author}
  {\bibfnamefont {Marc J.~J.}\ \bibnamefont {Vrakking}}, \bibinfo {author}
  {\bibfnamefont {Georg}\ \bibnamefont {Weidenspointner}}, \bibinfo {author}
  {\bibfnamefont {Thomas~A.}\ \bibnamefont {White}}, \bibinfo {author}
  {\bibfnamefont {Cornelia}\ \bibnamefont {Wunderer}}, \bibinfo {author}
  {\bibfnamefont {Gerard}\ \bibnamefont {Meijer}}, \bibinfo {author}
  {\bibfnamefont {Joachim}\ \bibnamefont {Ullrich}}, \bibinfo {author}
  {\bibfnamefont {Henrik}\ \bibnamefont {Stapelfeldt}}, \bibinfo {author}
  {\bibfnamefont {Daniel}\ \bibnamefont {Rolles}}, \ and\ \bibinfo {author}
  {\bibfnamefont {Henry~N.}\ \bibnamefont {Chapman}},\ }\bibfield  {title}
  {\enquote {\bibinfo {title} {Coherent diffractive imaging of controlled
  ensembles of isolated gas-phase molecules},}\ }\href@noop {} {\  (\bibinfo
  {year} {2013})},\ \bibinfo {note} {submitted},\ \Eprint
  {http://arxiv.org/abs/1307.4577} {arXiv:1307.4577 [physics]} \BibitemShut
  {NoStop}%
\bibitem [{\citenamefont {Block}\ \emph {et~al.}(1992)\citenamefont {Block},
  \citenamefont {Bohac},\ and\ \citenamefont {Miller}}]{Block:PRL68:1303}%
  \BibitemOpen
  \bibfield  {author} {\bibinfo {author} {\bibfnamefont {P.~A.}\ \bibnamefont
  {Block}}, \bibinfo {author} {\bibfnamefont {E.~J.}\ \bibnamefont {Bohac}}, \
  and\ \bibinfo {author} {\bibfnamefont {R.~E.}\ \bibnamefont {Miller}},\
  }\bibfield  {title} {\enquote {\bibinfo {title} {Spectroscopy of pendular
  states -- the use of molecular complexes in achieving orientation},}\ }\href
  {\doibase 10.1103/PhysRevLett.68.1303} {\bibfield  {journal} {\bibinfo
  {journal} {Phys.\ Rev.\ Lett.}\ }\textbf {\bibinfo {volume} {68}},\ \bibinfo
  {pages} {1303--1306} (\bibinfo {year} {1992})}\BibitemShut {NoStop}%
\bibitem [{\citenamefont {Kim}\ and\ \citenamefont
  {Felker}(1996)}]{Kim:JCP104:1147}%
  \BibitemOpen
  \bibfield  {author} {\bibinfo {author} {\bibfnamefont {W.}~\bibnamefont
  {Kim}}\ and\ \bibinfo {author} {\bibfnamefont {P.~M.}\ \bibnamefont
  {Felker}},\ }\bibfield  {title} {\enquote {\bibinfo {title} {Spectroscopy of
  pendular states in optical-field-aligned species},}\ }\href {\doibase
  10.1063/1.470770} {\bibfield  {journal} {\bibinfo  {journal} {J.\ Chem.\
  Phys.}\ }\textbf {\bibinfo {volume} {104}},\ \bibinfo {pages} {1147--1150}
  (\bibinfo {year} {1996})}\BibitemShut {NoStop}%
\bibitem [{\citenamefont {Stapelfeldt}\ \emph {et~al.}(1995)\citenamefont
  {Stapelfeldt}, \citenamefont {Constant},\ and\ \citenamefont
  {Corkum}}]{Stapelfeldt:PRL74:3780}%
  \BibitemOpen
  \bibfield  {author} {\bibinfo {author} {\bibfnamefont {Henrik}\ \bibnamefont
  {Stapelfeldt}}, \bibinfo {author} {\bibfnamefont {E}~\bibnamefont
  {Constant}}, \ and\ \bibinfo {author} {\bibfnamefont {P~B}\ \bibnamefont
  {Corkum}},\ }\bibfield  {title} {\enquote {\bibinfo {title} {Wave-packet
  structure and dynamics measured by coulomb explosion},}\ }\href {\doibase
  10.1103/PhysRevA.58.426} {\bibfield  {journal} {\bibinfo  {journal} {Phys.\
  Rev.\ Lett.}\ }\textbf {\bibinfo {volume} {74}},\ \bibinfo {pages}
  {3780--3783} (\bibinfo {year} {1995})}\BibitemShut {NoStop}%
\bibitem [{\citenamefont {Ortigoso}\ \emph {et~al.}(1999)\citenamefont
  {Ortigoso}, \citenamefont {Rodriguez}, \citenamefont {Gupta},\ and\
  \citenamefont {Friedrich}}]{Ortigoso:JCP110:3870}%
  \BibitemOpen
  \bibfield  {author} {\bibinfo {author} {\bibfnamefont {Juan}\ \bibnamefont
  {Ortigoso}}, \bibinfo {author} {\bibfnamefont {Mirta}\ \bibnamefont
  {Rodriguez}}, \bibinfo {author} {\bibfnamefont {Manish}\ \bibnamefont
  {Gupta}}, \ and\ \bibinfo {author} {\bibfnamefont {Bretislav}\ \bibnamefont
  {Friedrich}},\ }\bibfield  {title} {\enquote {\bibinfo {title} {Time
  evolution of pendular states created by the interaction of molecular
  polarizability with a pulsed nonresonant laser field},}\ }\href
  {http://link.aip.org/link/?JCP/110/3870/1} {\bibfield  {journal} {\bibinfo
  {journal} {J.\ Chem.\ Phys.}\ }\textbf {\bibinfo {volume} {110}},\ \bibinfo
  {pages} {3870--3875} (\bibinfo {year} {1999})}\BibitemShut {NoStop}%
\bibitem [{\citenamefont {Seideman}(1999)}]{Seideman:PRL83:4971}%
  \BibitemOpen
  \bibfield  {author} {\bibinfo {author} {\bibfnamefont {T}~\bibnamefont
  {Seideman}},\ }\bibfield  {title} {\enquote {\bibinfo {title} {Revival
  structure of aligned rotational wave packets},}\ }\href@noop {} {\bibfield
  {journal} {\bibinfo  {journal} {Phys.\ Rev.\ Lett.}\ }\textbf {\bibinfo
  {volume} {83}},\ \bibinfo {pages} {4971--4974} (\bibinfo {year}
  {1999})}\BibitemShut {NoStop}%
\bibitem [{\citenamefont {Torres}\ \emph {et~al.}(2005)\citenamefont {Torres},
  \citenamefont {de~Nalda},\ and\ \citenamefont
  {Marangos}}]{Torres:PRA72:023420}%
  \BibitemOpen
  \bibfield  {author} {\bibinfo {author} {\bibfnamefont {R}~\bibnamefont
  {Torres}}, \bibinfo {author} {\bibfnamefont {R}~\bibnamefont {de~Nalda}}, \
  and\ \bibinfo {author} {\bibfnamefont {J~P}\ \bibnamefont {Marangos}},\
  }\bibfield  {title} {\enquote {\bibinfo {title} {Dynamics of laser-induced
  molecular alignment in the impulsive and adiabatic regimes: A direct
  comparison},}\ }\href@noop {} {\bibfield  {journal} {\bibinfo  {journal}
  {Phys.\ Rev.\ A}\ }\textbf {\bibinfo {volume} {72}},\ \bibinfo {pages}
  {023420} (\bibinfo {year} {2005})}\BibitemShut {NoStop}%
\bibitem [{\citenamefont {Irimia}\ \emph {et~al.}(2009)\citenamefont {Irimia},
  \citenamefont {Dobrikov}, \citenamefont {Kortekaas}, \citenamefont {Voet},
  \citenamefont {van~den Ende}, \citenamefont {Groen},\ and\ \citenamefont
  {Janssen}}]{Irimia:RSI80:3958}%
  \BibitemOpen
  \bibfield  {author} {\bibinfo {author} {\bibfnamefont {Daniel}\ \bibnamefont
  {Irimia}}, \bibinfo {author} {\bibfnamefont {Dimitar}\ \bibnamefont
  {Dobrikov}}, \bibinfo {author} {\bibfnamefont {Rob}\ \bibnamefont
  {Kortekaas}}, \bibinfo {author} {\bibfnamefont {Han}\ \bibnamefont {Voet}},
  \bibinfo {author} {\bibfnamefont {Daan~A.}\ \bibnamefont {van~den Ende}},
  \bibinfo {author} {\bibfnamefont {Wilhelm~A.}\ \bibnamefont {Groen}}, \ and\
  \bibinfo {author} {\bibfnamefont {Maurice H.~M.}\ \bibnamefont {Janssen}},\
  }\bibfield  {title} {\enquote {\bibinfo {title} {A short pulse (7~$\mu$s
  fwhm) and high repetition rate (dc--5khz) cantilever piezovalve for pulsed
  atomic and molecular beams},}\ }\href {\doibase 10.1063/1.3263912} {\bibfield
   {journal} {\bibinfo  {journal} {Rev.\ Sci.\ Instrum.}\ }\textbf {\bibinfo
  {volume} {80}} (\bibinfo {year} {2009}),\ 10.1063/1.3263912}\BibitemShut
  {NoStop}%
\bibitem [{\citenamefont {Filsinger}\ \emph {et~al.}(2009)\citenamefont
  {Filsinger}, \citenamefont {K\"upper}, \citenamefont {Meijer}, \citenamefont
  {Holmegaard}, \citenamefont {Nielsen}, \citenamefont {Nevo}, \citenamefont
  {Hansen},\ and\ \citenamefont {Stapelfeldt}}]{Filsinger:JCP131:064309}%
  \BibitemOpen
  \bibfield  {author} {\bibinfo {author} {\bibfnamefont {Frank}\ \bibnamefont
  {Filsinger}}, \bibinfo {author} {\bibfnamefont {Jochen}\ \bibnamefont
  {K\"upper}}, \bibinfo {author} {\bibfnamefont {Gerard}\ \bibnamefont
  {Meijer}}, \bibinfo {author} {\bibfnamefont {Lotte}\ \bibnamefont
  {Holmegaard}}, \bibinfo {author} {\bibfnamefont {Jens~H.}\ \bibnamefont
  {Nielsen}}, \bibinfo {author} {\bibfnamefont {Iftach}\ \bibnamefont {Nevo}},
  \bibinfo {author} {\bibfnamefont {Jonas~L.}\ \bibnamefont {Hansen}}, \ and\
  \bibinfo {author} {\bibfnamefont {Henrik}\ \bibnamefont {Stapelfeldt}},\
  }\bibfield  {title} {\enquote {\bibinfo {title} {Quantum-state selection,
  alignment, and orientation of large molecules using static electric and laser
  fields},}\ }\href {\doibase 10.1063/1.3194287} {\bibfield  {journal}
  {\bibinfo  {journal} {J.\ Chem.\ Phys.}\ }\textbf {\bibinfo {volume} {131}},\
  \bibinfo {pages} {064309} (\bibinfo {year} {2009})},\ \Eprint
  {http://arxiv.org/abs/0903.5413} {arXiv:0903.5413 [physics]} \BibitemShut
  {NoStop}%
\bibitem [{\citenamefont {Nielsen}\ \emph {et~al.}(2011)\citenamefont
  {Nielsen}, \citenamefont {Simesen}, \citenamefont {Bisgaard}, \citenamefont
  {Stapelfeldt}, \citenamefont {Filsinger}, \citenamefont {Friedrich},
  \citenamefont {Meijer},\ and\ \citenamefont
  {K\"upper}}]{Nielsen:PCCP13:18971}%
  \BibitemOpen
  \bibfield  {author} {\bibinfo {author} {\bibfnamefont {Jens~H.}\ \bibnamefont
  {Nielsen}}, \bibinfo {author} {\bibfnamefont {Paw}\ \bibnamefont {Simesen}},
  \bibinfo {author} {\bibfnamefont {Christer~Z.}\ \bibnamefont {Bisgaard}},
  \bibinfo {author} {\bibfnamefont {Henrik}\ \bibnamefont {Stapelfeldt}},
  \bibinfo {author} {\bibfnamefont {Frank}\ \bibnamefont {Filsinger}}, \bibinfo
  {author} {\bibfnamefont {Bretislav}\ \bibnamefont {Friedrich}}, \bibinfo
  {author} {\bibfnamefont {Gerard}\ \bibnamefont {Meijer}}, \ and\ \bibinfo
  {author} {\bibfnamefont {Jochen}\ \bibnamefont {K\"upper}},\ }\bibfield
  {title} {\enquote {\bibinfo {title} {Stark-selected beam of ground-state
  {OCS} molecules characterized by revivals of impulsive alignment},}\ }\href
  {\doibase 10.1039/c1cp21143a} {\bibfield  {journal} {\bibinfo  {journal}
  {Phys.\ Chem.\ Chem.\ Phys.}\ }\textbf {\bibinfo {volume} {13}},\ \bibinfo
  {pages} {18971--18975} (\bibinfo {year} {2011})},\ \Eprint
  {http://arxiv.org/abs/1105.2413} {arXiv:1105.2413 [physics]} \BibitemShut
  {NoStop}%
\bibitem [{\citenamefont {Eppink}\ and\ \citenamefont
  {Parker}(1997)}]{Eppink:RSI68:3477}%
  \BibitemOpen
  \bibfield  {author} {\bibinfo {author} {\bibfnamefont {Andr\'{e} T. J.~B.}\
  \bibnamefont {Eppink}}\ and\ \bibinfo {author} {\bibfnamefont {David~H.}\
  \bibnamefont {Parker}},\ }\bibfield  {title} {\enquote {\bibinfo {title}
  {Velocity map imaging of ions and electrons using electrostatic lenses:
  Application in photoelectron and photofragment ion imaging of molecular
  oxygen},}\ }\href {\doibase 10.1063/1.1148310} {\bibfield  {journal}
  {\bibinfo  {journal} {Rev.\ Sci.\ Instrum.}\ }\textbf {\bibinfo {volume}
  {68}},\ \bibinfo {pages} {3477--3484} (\bibinfo {year} {1997})}\BibitemShut
  {NoStop}%
\bibitem [{\citenamefont {Trippel}\ \emph {et~al.}(2013)\citenamefont
  {Trippel}, \citenamefont {Mullins}, \citenamefont {Müller}, \citenamefont
  {Kienitz}, \citenamefont {D{\l}ugo{\l}{\k e}cki},\ and\ \citenamefont
  {K{\"u}pper}}]{Trippel:MP111:1738}%
  \BibitemOpen
  \bibfield  {author} {\bibinfo {author} {\bibfnamefont {Sebastian}\
  \bibnamefont {Trippel}}, \bibinfo {author} {\bibfnamefont {Terry}\
  \bibnamefont {Mullins}}, \bibinfo {author} {\bibfnamefont {Nele L.~M.}\
  \bibnamefont {Müller}}, \bibinfo {author} {\bibfnamefont {Jens~S.}\
  \bibnamefont {Kienitz}}, \bibinfo {author} {\bibfnamefont {Karol}\
  \bibnamefont {D{\l}ugo{\l}{\k e}cki}}, \ and\ \bibinfo {author}
  {\bibfnamefont {Jochen}\ \bibnamefont {K{\"u}pper}},\ }\bibfield  {title}
  {\enquote {\bibinfo {title} {Strongly aligned and oriented molecular samples
  at a khz repetition rate},}\ }\href@noop {} {\bibfield  {journal} {\bibinfo
  {journal} {Mol.\ Phys.}\ }\textbf {\bibinfo {volume} {111}},\ \bibinfo
  {pages} {1738} (\bibinfo {year} {2013})}\BibitemShut {NoStop}%
\bibitem [{\citenamefont {Sanchez-Moreno}\ \emph {et~al.}(2007)\citenamefont
  {Sanchez-Moreno}, \citenamefont {Gonz{\'a}lez-F{\'e}rez},\ and\ \citenamefont
  {Schmelcher}}]{SanchezMoreno:PRA76:053413}%
  \BibitemOpen
  \bibfield  {author} {\bibinfo {author} {\bibfnamefont {Pablo}\ \bibnamefont
  {Sanchez-Moreno}}, \bibinfo {author} {\bibfnamefont {Rosario}\ \bibnamefont
  {Gonz{\'a}lez-F{\'e}rez}}, \ and\ \bibinfo {author} {\bibfnamefont {Peter}\
  \bibnamefont {Schmelcher}},\ }\bibfield  {title} {\enquote {\bibinfo {title}
  {Molecular rotational dynamics in nonadiabatically switching homogeneous
  electric fields},}\ }\href {\doibase 10.1103/PhysRevA.76.053413} {\bibfield
  {journal} {\bibinfo  {journal} {Phys.\ Rev.\ A}\ }\textbf {\bibinfo {volume}
  {76}},\ \bibinfo {pages} {053413} (\bibinfo {year} {2007})}\BibitemShut
  {NoStop}%
\bibitem [{\citenamefont {Omiste}\ \emph {et~al.}(2011)\citenamefont {Omiste},
  \citenamefont {Gaerttner}, \citenamefont {Schmelcher}, \citenamefont
  {Gonz{\'a}lez-F{\'e}rez}, \citenamefont {Holmegaard}, \citenamefont
  {Nielsen}, \citenamefont {Stapelfeldt},\ and\ \citenamefont
  {K{\"u}pper}}]{Omiste:PCCP13:18815}%
  \BibitemOpen
  \bibfield  {author} {\bibinfo {author} {\bibfnamefont {Juan~J.}\ \bibnamefont
  {Omiste}}, \bibinfo {author} {\bibfnamefont {M.}~\bibnamefont {Gaerttner}},
  \bibinfo {author} {\bibfnamefont {P.}~\bibnamefont {Schmelcher}}, \bibinfo
  {author} {\bibfnamefont {R.}~\bibnamefont {Gonz{\'a}lez-F{\'e}rez}}, \bibinfo
  {author} {\bibfnamefont {Lotte}\ \bibnamefont {Holmegaard}}, \bibinfo
  {author} {\bibfnamefont {Jens~Hedegaard}\ \bibnamefont {Nielsen}}, \bibinfo
  {author} {\bibfnamefont {Henrik}\ \bibnamefont {Stapelfeldt}}, \ and\
  \bibinfo {author} {\bibfnamefont {Jochen}\ \bibnamefont {K{\"u}pper}},\
  }\bibfield  {title} {\enquote {\bibinfo {title} {Theoretical description of
  adiabatic laser alignment and mixed-field orientation: the need for a
  non-adiabatic model},}\ }\href {\doibase 10.1039/c1cp21195a} {\bibfield
  {journal} {\bibinfo  {journal} {Phys.\ Chem.\ Chem.\ Phys.}\ }\textbf
  {\bibinfo {volume} {13}},\ \bibinfo {pages} {18815--18824} (\bibinfo {year}
  {2011})},\ \Eprint {http://arxiv.org/abs/1105.0534} {arXiv:1105.0534
  [physics]} \BibitemShut {NoStop}%
\bibitem [{\citenamefont {Lee}\ \emph {et~al.}(2004)\citenamefont {Lee},
  \citenamefont {Villeneuve}, \citenamefont {Corkum},\ and\ \citenamefont
  {Shapiro}}]{Lee:PRL93:233601}%
  \BibitemOpen
  \bibfield  {author} {\bibinfo {author} {\bibfnamefont {K.~F.}\ \bibnamefont
  {Lee}}, \bibinfo {author} {\bibfnamefont {D.~M.}\ \bibnamefont {Villeneuve}},
  \bibinfo {author} {\bibfnamefont {P.~B.}\ \bibnamefont {Corkum}}, \ and\
  \bibinfo {author} {\bibfnamefont {E.~A.}\ \bibnamefont {Shapiro}},\
  }\bibfield  {title} {\enquote {\bibinfo {title} {Phase control of rotational
  wave packets and quantum information},}\ }\href@noop {} {\bibfield  {journal}
  {\bibinfo  {journal} {Phys.\ Rev.\ Lett.}\ }\textbf {\bibinfo {volume}
  {93}},\ \bibinfo {pages} {233601} (\bibinfo {year} {2004})}\BibitemShut
  {NoStop}%
\bibitem [{\citenamefont {Filsinger}\ \emph {et~al.}(2011)\citenamefont
  {Filsinger}, \citenamefont {Meijer}, \citenamefont {Stapelfeldt},
  \citenamefont {Chapman},\ and\ \citenamefont
  {K\"upper}}]{Filsinger:PCCP13:2076}%
  \BibitemOpen
  \bibfield  {author} {\bibinfo {author} {\bibfnamefont {Frank}\ \bibnamefont
  {Filsinger}}, \bibinfo {author} {\bibfnamefont {Gerard}\ \bibnamefont
  {Meijer}}, \bibinfo {author} {\bibfnamefont {Henrik}\ \bibnamefont
  {Stapelfeldt}}, \bibinfo {author} {\bibfnamefont {Henry}\ \bibnamefont
  {Chapman}}, \ and\ \bibinfo {author} {\bibfnamefont {Jochen}\ \bibnamefont
  {K\"upper}},\ }\bibfield  {title} {\enquote {\bibinfo {title} {State- and
  conformer-selected beams of aligned and oriented molecules for ultrafast
  diffraction studies},}\ }\href {\doibase 10.1039/C0CP01585G} {\bibfield
  {journal} {\bibinfo  {journal} {Phys.\ Chem.\ Chem.\ Phys.}\ }\textbf
  {\bibinfo {volume} {13}},\ \bibinfo {pages} {2076--2087} (\bibinfo {year}
  {2011})}\BibitemShut {NoStop}%
\bibitem [{\citenamefont {Hensley}\ \emph {et~al.}(2012)\citenamefont
  {Hensley}, \citenamefont {Yang},\ and\ \citenamefont
  {Centurion}}]{Hensley:PRL109:133202}%
  \BibitemOpen
  \bibfield  {author} {\bibinfo {author} {\bibfnamefont {Christopher~J.}\
  \bibnamefont {Hensley}}, \bibinfo {author} {\bibfnamefont {Jie}\ \bibnamefont
  {Yang}}, \ and\ \bibinfo {author} {\bibfnamefont {Martin}\ \bibnamefont
  {Centurion}},\ }\bibfield  {title} {\enquote {\bibinfo {title} {Imaging of
  isolated molecules with ultrafast electron pulses},}\ }\href {\doibase
  10.1103/PhysRevLett.109.133202} {\bibfield  {journal} {\bibinfo  {journal}
  {Phys.\ Rev.\ Lett.}\ }\textbf {\bibinfo {volume} {109}},\ \bibinfo {pages}
  {133202} (\bibinfo {year} {2012})}\BibitemShut {NoStop}%
\bibitem [{\citenamefont {Choi}\ \emph {et~al.}(2006)\citenamefont {Choi},
  \citenamefont {Kahng}, \citenamefont {Kim}, \citenamefont {Kim},
  \citenamefont {Kim}, \citenamefont {Song}, \citenamefont {Ihm},\ and\
  \citenamefont {Kuk}}]{Choi:PRL96:156106}%
  \BibitemOpen
  \bibfield  {author} {\bibinfo {author} {\bibfnamefont {Byoung-Young}\
  \bibnamefont {Choi}}, \bibinfo {author} {\bibfnamefont {Se-Jong}\
  \bibnamefont {Kahng}}, \bibinfo {author} {\bibfnamefont {Seungchul}\
  \bibnamefont {Kim}}, \bibinfo {author} {\bibfnamefont {Hajin}\ \bibnamefont
  {Kim}}, \bibinfo {author} {\bibfnamefont {Hyo}\ \bibnamefont {Kim}}, \bibinfo
  {author} {\bibfnamefont {Young}\ \bibnamefont {Song}}, \bibinfo {author}
  {\bibfnamefont {Jisoon}\ \bibnamefont {Ihm}}, \ and\ \bibinfo {author}
  {\bibfnamefont {Young}\ \bibnamefont {Kuk}},\ }\bibfield  {title} {\enquote
  {\bibinfo {title} {Conformational molecular switch of the azobenzene
  molecule: A scanning tunneling microscopy study},}\ }\href {\doibase
  10.1103/PhysRevLett.96.156106} {\bibfield  {journal} {\bibinfo  {journal}
  {Phys.\ Rev.\ Lett.}\ }\textbf {\bibinfo {volume} {96}},\ \bibinfo {pages}
  {156106} (\bibinfo {year} {2006})}\BibitemShut {NoStop}%
\bibitem [{\citenamefont {Avellini}\ \emph {et~al.}(2012)\citenamefont
  {Avellini}, \citenamefont {Li}, \citenamefont {Coskun}, \citenamefont
  {Barin}, \citenamefont {Trabolsi}, \citenamefont {Basuray}, \citenamefont
  {Dey}, \citenamefont {Credi}, \citenamefont {Silvi}, \citenamefont
  {Stoddart},\ and\ \citenamefont {Venturi}}]{Avellini:ACIE51:1611}%
  \BibitemOpen
  \bibfield  {author} {\bibinfo {author} {\bibfnamefont {Tommaso}\ \bibnamefont
  {Avellini}}, \bibinfo {author} {\bibfnamefont {Hao}\ \bibnamefont {Li}},
  \bibinfo {author} {\bibfnamefont {Ali}\ \bibnamefont {Coskun}}, \bibinfo
  {author} {\bibfnamefont {Gokhan}\ \bibnamefont {Barin}}, \bibinfo {author}
  {\bibfnamefont {Ali}\ \bibnamefont {Trabolsi}}, \bibinfo {author}
  {\bibfnamefont {Ashish~N}\ \bibnamefont {Basuray}}, \bibinfo {author}
  {\bibfnamefont {Sanjeev~K}\ \bibnamefont {Dey}}, \bibinfo {author}
  {\bibfnamefont {Alberto}\ \bibnamefont {Credi}}, \bibinfo {author}
  {\bibfnamefont {Serena}\ \bibnamefont {Silvi}}, \bibinfo {author}
  {\bibfnamefont {J~Fraser}\ \bibnamefont {Stoddart}}, \ and\ \bibinfo {author}
  {\bibfnamefont {Margherita}\ \bibnamefont {Venturi}},\ }\bibfield  {title}
  {\enquote {\bibinfo {title} {Photoinduced memory effect in a redox
  controllable bistable mechanical molecular switch},}\ }\href {\doibase
  10.1002/anie.201107618} {\bibfield  {journal} {\bibinfo  {journal} {Angew.\
  Chem.\ Int.\ Ed.}\ }\textbf {\bibinfo {volume} {51}},\ \bibinfo {pages}
  {1611--1615} (\bibinfo {year} {2012})}\BibitemShut {NoStop}%
\bibitem [{\citenamefont {Ramakrishna}\ and\ \citenamefont
  {Seideman}(2007)}]{Ramakrishna:PRL99:103001}%
  \BibitemOpen
  \bibfield  {author} {\bibinfo {author} {\bibfnamefont {S.}~\bibnamefont
  {Ramakrishna}}\ and\ \bibinfo {author} {\bibfnamefont {T.}~\bibnamefont
  {Seideman}},\ }\bibfield  {title} {\enquote {\bibinfo {title} {Torsional
  control by intense pulses},}\ }\href {\doibase 10.1103/PhysRevLett.99.103001}
  {\bibfield  {journal} {\bibinfo  {journal} {Phys.\ Rev.\ Lett.}\ }\textbf
  {\bibinfo {volume} {99}},\ \bibinfo {pages} {103001} (\bibinfo {year}
  {2007})}\BibitemShut {NoStop}%
\bibitem [{\citenamefont {Donhauser}\ \emph {et~al.}(2001)\citenamefont
  {Donhauser}, \citenamefont {Mantooth}, \citenamefont {Kelly}, \citenamefont
  {Bumm}, \citenamefont {Monnell}, \citenamefont {Stapleton}, \citenamefont
  {Price}, \citenamefont {Rawlett}, \citenamefont {Allara}, \citenamefont
  {Tour},\ and\ \citenamefont {Weiss}}]{Donhauser:Science292:2303}%
  \BibitemOpen
  \bibfield  {author} {\bibinfo {author} {\bibfnamefont {Z~J}\ \bibnamefont
  {Donhauser}}, \bibinfo {author} {\bibfnamefont {B~A}\ \bibnamefont
  {Mantooth}}, \bibinfo {author} {\bibfnamefont {K~F}\ \bibnamefont {Kelly}},
  \bibinfo {author} {\bibfnamefont {L~A}\ \bibnamefont {Bumm}}, \bibinfo
  {author} {\bibfnamefont {J~D}\ \bibnamefont {Monnell}}, \bibinfo {author}
  {\bibfnamefont {J~J}\ \bibnamefont {Stapleton}}, \bibinfo {author}
  {\bibfnamefont {D~W}\ \bibnamefont {Price}}, \bibinfo {author} {\bibfnamefont
  {A~M}\ \bibnamefont {Rawlett}}, \bibinfo {author} {\bibfnamefont {D~L}\
  \bibnamefont {Allara}}, \bibinfo {author} {\bibfnamefont {J~M}\ \bibnamefont
  {Tour}}, \ and\ \bibinfo {author} {\bibfnamefont {P~S}\ \bibnamefont
  {Weiss}},\ }\bibfield  {title} {\enquote {\bibinfo {title} {Conductance
  switching in single molecules through conformational changes},}\ }\href
  {\doibase 10.1126/science.1060294} {\bibfield  {journal} {\bibinfo  {journal}
  {Science}\ }\textbf {\bibinfo {volume} {292}},\ \bibinfo {pages} {2303--2307}
  (\bibinfo {year} {2001})}\BibitemShut {NoStop}%
\bibitem [{\citenamefont {Reuter}\ \emph {et~al.}(2008)\citenamefont {Reuter},
  \citenamefont {Sukharev},\ and\ \citenamefont
  {Seideman}}]{Reuter:PRL101:208303}%
  \BibitemOpen
  \bibfield  {author} {\bibinfo {author} {\bibfnamefont {M.~G.}\ \bibnamefont
  {Reuter}}, \bibinfo {author} {\bibfnamefont {M.}~\bibnamefont {Sukharev}}, \
  and\ \bibinfo {author} {\bibfnamefont {T.}~\bibnamefont {Seideman}},\
  }\bibfield  {title} {\enquote {\bibinfo {title} {Laser field alignment of
  organic molecules on semiconductor surfaces: Toward ultrafast molecular
  switches},}\ }\href {\doibase 10.1103/PhysRevLett.101.208303} {\bibfield
  {journal} {\bibinfo  {journal} {Phys.\ Rev.\ Lett.}\ }\textbf {\bibinfo
  {volume} {101}},\ \bibinfo {pages} {208303} (\bibinfo {year}
  {2008})}\BibitemShut {NoStop}%
\bibitem [{\citenamefont {Gordy}\ and\ \citenamefont
  {Cook}(1984)}]{Gordy:MWMolSpec}%
  \BibitemOpen
  \bibfield  {author} {\bibinfo {author} {\bibfnamefont {W.}~\bibnamefont
  {Gordy}}\ and\ \bibinfo {author} {\bibfnamefont {R.~L.}\ \bibnamefont
  {Cook}},\ }\href@noop {} {\emph {\bibinfo {title} {Microwave Molecular
  Spectra}}},\ \bibinfo {edition} {3rd}\ ed.\ (\bibinfo  {publisher} {John
  Wiley \& Sons},\ \bibinfo {address} {New York, NY, USA},\ \bibinfo {year}
  {1984})\BibitemShut {NoStop}%
\bibitem [{\citenamefont {del Valle}\ \emph {et~al.}(2007)\citenamefont {del
  Valle}, \citenamefont {Guti\'errez}, \citenamefont {Tejedor},\ and\
  \citenamefont {Cuniberti}}]{delValle:NatNano2:176}%
  \BibitemOpen
  \bibfield  {author} {\bibinfo {author} {\bibfnamefont {Miriam}\ \bibnamefont
  {del Valle}}, \bibinfo {author} {\bibfnamefont {Rafael}\ \bibnamefont
  {Guti\'errez}}, \bibinfo {author} {\bibfnamefont {Carlos}\ \bibnamefont
  {Tejedor}}, \ and\ \bibinfo {author} {\bibfnamefont {Gianaurelio}\
  \bibnamefont {Cuniberti}},\ }\bibfield  {title} {\enquote {\bibinfo {title}
  {Tuning the conductance of a molecular switch},}\ }\href {\doibase
  10.1038/nnano.2007.38} {\bibfield  {journal} {\bibinfo  {journal} {Nat.
  Nanotechnol.}\ }\textbf {\bibinfo {volume} {2}},\ \bibinfo {pages} {176--179}
  (\bibinfo {year} {2007})}\BibitemShut {NoStop}%
\bibitem [{\citenamefont {Gro{\ss}mann}\ \emph {et~al.}(2002)\citenamefont
  {Gro{\ss}mann}, \citenamefont {Feng}, \citenamefont {Schmidt}, \citenamefont
  {Kunert},\ and\ \citenamefont {Schmidt}}]{Grossmann:EPL60:201}%
  \BibitemOpen
  \bibfield  {author} {\bibinfo {author} {\bibfnamefont {F}~\bibnamefont
  {Gro{\ss}mann}}, \bibinfo {author} {\bibfnamefont {L}~\bibnamefont {Feng}},
  \bibinfo {author} {\bibfnamefont {G}~\bibnamefont {Schmidt}}, \bibinfo
  {author} {\bibfnamefont {T}~\bibnamefont {Kunert}}, \ and\ \bibinfo {author}
  {\bibfnamefont {R}~\bibnamefont {Schmidt}},\ }\bibfield  {title} {\enquote
  {\bibinfo {title} {Optimal control of a molecular cis-trans isomerization
  model},}\ }\href {\doibase 10.1209/epl/i2002-00339-6} {\bibfield  {journal}
  {\bibinfo  {journal} {Europhys.\ Lett.}\ }\textbf {\bibinfo {volume} {60}},\
  \bibinfo {pages} {201--206} (\bibinfo {year} {2002})}\BibitemShut {NoStop}%
\bibitem [{\citenamefont {Barbatti}\ \emph {et~al.}(2008)\citenamefont
  {Barbatti}, \citenamefont {Belz}, \citenamefont {Leibscher}, \citenamefont
  {Lischka},\ and\ \citenamefont {Manz}}]{Barbatti:CP350:145}%
  \BibitemOpen
  \bibfield  {author} {\bibinfo {author} {\bibfnamefont {M}~\bibnamefont
  {Barbatti}}, \bibinfo {author} {\bibfnamefont {S}~\bibnamefont {Belz}},
  \bibinfo {author} {\bibfnamefont {M}~\bibnamefont {Leibscher}}, \bibinfo
  {author} {\bibfnamefont {H}~\bibnamefont {Lischka}}, \ and\ \bibinfo {author}
  {\bibfnamefont {J}~\bibnamefont {Manz}},\ }\bibfield  {title} {\enquote
  {\bibinfo {title} {Sensitivity of femtosecond quantum dynamics and control
  with respect to non-adiabatic couplings: Model simulations for the cis--trans
  isomerization of the dideuterated methaniminium cation},}\ }\href {\doibase
  10.1016/j.chemphys.2008.01.053} {\bibfield  {journal} {\bibinfo  {journal}
  {Chem.\ Phys.}\ }\textbf {\bibinfo {volume} {350}},\ \bibinfo {pages}
  {145--153} (\bibinfo {year} {2008})}\BibitemShut {NoStop}%
\bibitem [{\citenamefont {Parker}\ \emph {et~al.}(2011)\citenamefont {Parker},
  \citenamefont {Ratner},\ and\ \citenamefont
  {Seideman}}]{Parker:JCP135:224301}%
  \BibitemOpen
  \bibfield  {author} {\bibinfo {author} {\bibfnamefont {Shane~M}\ \bibnamefont
  {Parker}}, \bibinfo {author} {\bibfnamefont {Mark~A}\ \bibnamefont {Ratner}},
  \ and\ \bibinfo {author} {\bibfnamefont {Tamar}\ \bibnamefont {Seideman}},\
  }\bibfield  {title} {\enquote {\bibinfo {title} {Coherent control of
  molecular torsion},}\ }\href {\doibase 10.1063/1.3663710} {\bibfield
  {journal} {\bibinfo  {journal} {J.\ Chem.\ Phys.}\ }\textbf {\bibinfo
  {volume} {135}},\ \bibinfo {pages} {224301} (\bibinfo {year}
  {2011})}\BibitemShut {NoStop}%
\end{thebibliography}%
\end{document}